\documentstyle[epsfig,referee]{mn}

\newcommand{\reference}{\bibitem}
\def\beq{\begin{equation}}
\def\eeq{\end{equation}}
\def\bey{\begin{eqnarray}}
\def\eey{\end{eqnarray}}
\def\beqarray{\begin{eqnarray}}
\def\eeqarray{\end{eqnarray}}

\def\v200{V_{200}}

\title[]{Gamma-ray bursts: afterglows from cylindrical jets} 

\author[]{K. S. Cheng,$^{1 \; \star}$ Y. F. Huang$^{2,3,4 \; \star}$ 
   and T. Lu$^{2,4}$ 
\thanks{E-mail: hrspksc@hkucc.hku.hk(KSC); hyf@nju.edu.cn(YFH);
        tlu@nju.edu.cn(TL)} \\
$^1${\sl Department of Physics, The University of Hong Kong, 
         Pokfulam Road, Hong Kong, China} \\
$^2${\sl Department of Astronomy, Nanjing University, Nanjing 210093, 
         China} \\
$^3${\sl Astronomical and Astrophysical Center of East China, 
         Nanjing University, Nanjing 210093, China} \\
$^4${\sl LCRHEA, Institute for High-Energy Physics, Chinese Academy of 
         Sciences, Beijing 100039, China} } 
\date{Accepted ........
      Received .......;
      in original form .......}
\pubyear{2001}
\begin{document}
\voffset=-0.5 in

\maketitle
\begin{abstract}
Nearly all previous discussion on beaming effects in gamma-ray bursts
have assumed a conical geometry. However, more and more observations on
relativistic jets in radio galaxies, active galactic nuclei, and
``microquasars'' in the Galaxy have shown that many of these outflows
are not conical, but cylindrical, i.e., they maintain constant cross
sections at large scales. Thus it is necessary to discuss the possibility
that gamma-ray bursts may be due to highly collimated cylindrical jets,
not conical ones. Here we study the dynamical evolution of cylindrical jets
and discuss their afterglows. Both analytical and numerical results are
presented. It is shown that when the lateral expansion is not taken into
account, a cylindrical jet typically remains to be highly relativistic 
for $\sim 10^8$ --- $10^9$ s. During this relativistic phase, the optical 
afterglow decays as $S_{\nu} \propto t^{-p/2}$ at first, where $p$ is the 
index characterizing the power-law energy distribution of electrons. 
Then the light curve steepens to be $S_{\nu} \propto t^{-(p+1)/2}$ due to 
cooling of electrons. After entering the non-relativistic phase (i.e., 
$t \geq 10^{11}$ s), the afterglow is $S_{\nu} \propto t^{-(5p-4)/6}$.
But if the cylindrical jet expands laterally at co-moving sound speed, 
then the decay becomes $S_{\nu} \propto t^{-p}$ 
and $S_{\nu} \propto t^{-(15p-21)/10}$ --- $t^{-(15p-20)/10}$ in the
ultra-relativistic and non-relativistic phase respectively. Note that
in both cases, the light curve turns flatter after the
relativistic-Newtonian transition point, which differs markedly from
the behaviour of a conical jet. It is suggested that some gamma-ray
bursts with afterglows decaying as $t^{-1.1}$ --- $t^{-1.3}$ may be
due to cylindrical jets, not necessarily isotropic fireballs.
\end{abstract}

\section {Introduction}

The discovery of gamma-ray burst (GRB) afterglows in early 1997 
(van Paradijs et al. 1997; Sahu et al. 1997) 
has revolutionized the researches in the field (Wijers 1998).
Since then, X-ray afterglows (Costa et al. 1997; Piro et al. 1998) have
been detected from more than 20 GRBs, of which a half were observed
optically (e.g., Groot et al. 1997; Galama et al. 1997; Metzger et al. 1997;
Kulkarni et al. 1998; Pedersen et al. 1998; Garcia et al. 1998; Akerlof et
al. 1999; Vreeswijk et al. 1999) and one third were even observed in 
radio bands (e.g., Frail et al. 1997; Galama et al. 1998a, b). 
The cosmological origin is thus firmly established, and
the so called fireball model (Goodman 1986; Paczy\'{n}ski 1986; Rees
\& M\'{e}sz\'{a}ros 1992, 1994; M\'{e}sz\'{a}ros \& Rees 1992;
M\'{e}sz\'{a}ros, Laguna \& Rees 1993; M\'{e}sz\'{a}ros, Rees \&
Papathanassiou 1994; Katz 1994; Sari, Narayan \& Piran 1996) becomes
the most popular model, which is found successful in explaining the
global features of GRB afterglows (M\'{e}sz\'{a}ros \& Rees 1997;
Waxman 1997a, b; Tavani 1997; Vietri 1997; Wijers, Rees \& M\'{e}sz\'{a}ros
1997; M\'{e}sz\'{a}ros, Rees \& Wijers 1998; Sari, Piran \& Narayan 
1998; Dermer et al. 1999a, b; Wijers \& Galama 1999; Dai \& Lu 1998,
1999). For detailed reviews on recent progresses, see Piran (1999) and
van Paradijs, Kouveliotou \& Wijers (2000).

However, we are still far from resolving the puzzle of GRBs, because
many crucial problems are largely open. For example, whether a GRB is
due to a highly collimated jet or an isotropic fireball is still
controversial. This issue has been discussed extensively. It is
generally believed that due to both the edge effect (Panaitescu \&
M\'{e}sz\'{a}ros 1999; Kulkarni et al. 1999; M\'{e}sz\'{a}ros \&
Rees 1999) and the lateral expansion effect (Rhoads 1997, 1999),
afterglows from a jet will be characterized by a break in the light
curve during the relativistic phase. GRBs 990123, 990510 and 000301c
are regarded as good examples (Kulkarni et al. 1999; Harrison et al.
1999; Castro-Tirado et al. 1999; Sari, Piran \& Halpern 1999;
Wijers et al. 1999; Rhoads \& Fruchter 2001; Berger et al. 2001; 
Jensen et al. 2001; Masetti et al. 2000). But detailed numerical
studies show that the break is usually quite smooth (Panaitescu \&
M\'{e}sz\'{a}ros 1998; Moderski, Sikora \& Bulik 2000). Recently
Huang et al. (1998a, b, 1999a, b) stressed the importance of the
non-relativistic phase of GRB afterglows. They have used their
refined dynamical model to discuss afterglows from jetted GRB remnants
(Huang et al. 2000a, b, c). It is found that the most 
obvious light curve break does
not appear in the relativistic phase, but occurs at the
relativistic-Newtonian transition point. They also stressed that the
break is parameter-dependent (Huang et al. 2000b). Based on their
investigations, the rapid fading of optical afterglows from GRBs 970228,
980326, 980519, 990123, 990510 and 991208 has been suggested as
evidence for beaming in these cases (Huang et al. 2000d).

In all these discussion, a conical jet (i.e., a jet with a constant
half opening angle; of course, in some more realistic discussion,
the conical jet is permitted to expand laterally) has been assumed.
However, we should note that this might not be the case. For
example, we know that a relativistic flow (i.e., a jet) is a common
feature of radio galaxies. Interestingly, copious examples have
shown that these outflows are not conical, instead, they maintain
constant cross sections at large scales (Perley, Bridle \& Willis
1984; Biretta, Sparks \& Macchetto 1999; also see GLAST home page:
{\sl http://www-glast.sonoma.edu/gru/agn/Default.htm} and Chandra home
page: {\sl http://chandra.harvard.edu/photo/cycle1/pictor/index.html}
for good examples). Here we call a jet maintaining a constant
cross section a cylindrical jet. It is reasonable to deduce that
ultra-relativistic outflows in GRBs might also be cylindrical jets,
not conical ones. In fact, this idea has already been suggested
as GRB trigger mechanism by Dar et al. (Shaviv \& Dar 1995; Dar
1998, 1999).

Afterglows from a cylindrical jet is likely to differ from those of a
conical jet markedly. If GRBs are really due to cylindrical jets, then
we need to examine their dynamics and afterglows in detail, which is
just the purpose of this article. The structure of our article is as
follows. We describe our model in Section 2 and present the analytic
solution in Section 3. Section 4 is the numerical results and Section
5 is brief discussion.

\section{Model}

Let us consider a highly collimated GRB remnant, ploughing its way
through a homogeneous interstellar medium (ISM) where the number
density is $n$. Denote the lateral radius of the remnant as $a$,
and its distance from the central engine of the GRB as $R$. The
remnant will maintain a constant cross section (i.e., $a \equiv$ const),
except that it may expand laterally at a speed $v_{\perp}$.

\subsection{Dynamics}

In the literature, the following differential equation is usually used
to depict the expansion of GRB remnants (e.g., Chiang \& Dermer 1999;
Piran 1999),
\begin{equation}
\label{odgm1}
\frac{d \gamma}{d m} = - \frac{\gamma^2 - 1} {M},
\end{equation}
where $m$ is the rest mass of the swept-up medium, $\gamma$ is the
bulk Lorentz factor and $M$ is the total mass in the co-moving frame,
including internal energy. However, Huang et al. (1999a, b) pointed
out that in the adiabatic case equation~(\ref{odgm1}) does not match
with the Sedov solution during the non-relativistic phase. They have
proposed a refined equation,
\begin{equation}
\label{ndgm2}
\frac{d \gamma}{d m} = - \frac{\gamma^2 - 1}
       {M_{\rm ej} + \epsilon m + 2 ( 1 - \epsilon) \gamma m}, 
\end{equation}
which is shown to be correct for both adiabatic and highly radiative
shocks, and in both relativistic and Newtonian phase. Here $M_{\rm ej}$
is the initial ejecta mass and $\epsilon$ is the radiative efficiency.
In this paper we adopt equation~(\ref{ndgm2}) to describe the
dynamical behavior of our cylindrical jet. For simplicity, we will only
consider the adiabatic case (i.e., $\epsilon \equiv 0$).

The evolution of radius ($R$), the swept-up mass ($m$), and the lateral
radius ($a$) is described by (cf. Huang et al. 2000a, b),
\begin{equation}
\label{drdt3}
\frac{d R}{d t} = \beta c \gamma (\gamma + \sqrt{\gamma^2 - 1}),
\end{equation}
\begin{equation}
\label{dmdr4}
\frac{d m}{d R} = \pi a^2 n m_{\rm p},
\end{equation}
\begin{equation}
\label{dadt5}
\frac{d a}{d t} = v_{\perp} (\gamma + \sqrt{\gamma^2 - 1}),
\end{equation}
where $t$ is the observer's time, $\beta = \sqrt{\gamma^2 - 1}/ \gamma$,
$c$ is the speed of light, and $m_{\rm p}$ is the mass of proton. As for
$v_{\perp}$, we have two possibilities, just like in the case of a
conical jet: (i) $v_{\perp} \equiv 0$, this means the jet does not expand
laterally, it maintains a constant cross section strictly. The problem is
then greatly simplified; (ii) $v_{\perp} \equiv c_{\rm s}$, i.e., the jet
expands laterally at the co-moving sound speed, of which a reasonable
expression has been derived by Huang et al. (2000a, b) as, 
\begin{equation}
\label{cs6}
c_{\rm s}^2 = \hat{\gamma} (\hat{\gamma} - 1) (\gamma - 1) 
              \frac{1}{1 + \hat{\gamma}(\gamma - 1)} c^2 ,
\end{equation}
where $\hat{\gamma} \approx (4 \gamma + 1)/(3 \gamma)$ is the adiabatic
index (Dai, Huang \& Lu 1999). 
In the ultra-relativistic limit ($\gamma \gg 1, \hat{\gamma} \approx 4/3$), 
equation~(\ref{cs6}) gives $c_{\rm s}^2 = c^2/3$; and in the 
non-relativistic limit ($\gamma \sim 1, \hat{\gamma} \approx 5/3$), we 
simply get $c_{\rm s}^2 = 5 \beta^2 c^2/9$.

Equations~(\ref{ndgm2}) --- (\ref{dadt5}) give a thorough description of
the dynamical evolution of our cylindrical jets. We will solve them
analytically in the next section and then present our numerical results
in Section 4.

\subsection{Synchrotron radiation}

A strong blastwave will be produced due to the interaction of the
remnant with the ISM. Synchrotron radiation from the shock-accelerated 
ISM electrons gives birth to GRB afterglows. As usual, we assume that 
the magnetic energy density in the co-moving frame is a 
fraction $\xi_{\rm B}^2$ of the total thermal energy density 
($B'^2 / 8 \pi = \xi_{\rm B}^2  e'$), and that electrons carry 
a fraction $\xi_{\rm e}$ of the proton energy, which means that the 
minimum Lorentz factor of the random motion of electrons in 
the co-moving frame is 
$\gamma_{\rm e,min} = \xi_{\rm e} (\gamma - 1) 
                     m_{\rm p} / m_{\rm e} + 1$, 
where $m_{\rm e}$ is the electron mass.
In the absence of radiation loss, the distribution of the 
shock-accelerated electrons behind the blastwave can be 
characterized by a power-law function of electron energy,
$d N_{\rm e}'/d \gamma_{\rm e} \propto \gamma_{\rm e}^{-p} $  
$(\gamma_{\rm e,min} \leq \gamma_{\rm e} \leq \gamma_{\rm e,max})$,
where $\gamma_{\rm e,max}$ is the maximum Lorentz factor,
$\gamma_{\rm e,max} = 10^8 (B' /1 {\rm G})^{-1/2}$ (M\'{e}sz\'{a}ros
et al. 1993; Vietri 1997; Dai et al. 1999). 

However, radiation loss may play an important role in the process. 
Sari et al. (1998) have derived an equation for the critical electron
Lorentz factor, $\gamma_{\rm c}$, above which synchrotron radiation 
is significant: 
\begin{equation}
\label{gammac7}
\gamma_{\rm c} = 6 \pi m_{\rm e} c / (\sigma_{\rm T} \gamma B'^2 t),
\end{equation}
where $\sigma_{\rm T}$ is the Thompson cross section.  
The distribution of radiative electrons (i.e., those with 
$\gamma_{\rm e} > \gamma_{\rm c}$) becomes another power-law function 
of $d N_{\rm e}'/d \gamma_{\rm e} \propto \gamma_{\rm e}^{-(p+1)} $.
A good description of electron distribution allowing for radiation loss 
has been presented by Dai et al. (1999, also see: Huang et al. 2000a, b). 
We will use their expression in this research. For convenience, let us 
define a characteristic frequency of 
$\nu_{\rm c} \equiv \gamma \gamma_{\rm c}^2 e B' / (2 \pi m_{\rm e} c)$ 
here (Sari et al. 1998). 

In the co-moving frame, synchrotron radiation power at frequency
$\nu '$ from electrons is given by (Rybicki \& Lightman 1979),
\begin{equation}
\label{pnu8}
P'(\nu ') = \frac{\sqrt{3} e^3 B'}{m_{\rm e} c^2} 
	    \int_{\gamma_{\rm e,min}}^{\gamma_{\rm e,max}} 
	    \left( \frac{dN_{\rm e}'}{d\gamma_{\rm e}} \right)
	    F\left(\frac{\nu '}{\nu_{\rm cr}'} \right) d\gamma_{\rm e},
\end{equation}
where $e$ is electron charge, 
$\nu_{\rm cr}' = 3 \gamma_{\rm e}^2 e B' / (4 \pi m_{\rm e} c)$, and 
\begin{equation}
\label{fx9}
F(x) = x \int_{x}^{+ \infty} K_{5/3}(k) dk,
\end{equation}
with $K_{5/3}(k)$ being the Bessel function. We assume that this power 
is radiated isotropically,
\begin{equation}
\label{dpdw10}
\frac{d P'(\nu ')}{d \Omega '} = \frac{P'(\nu ')}{4 \pi}.
\end{equation}

Let $\Theta$ be the angle between the axis of the jet and the line of
sight, and define $\mu = \cos \Theta$, we can derive the angular
distribution of power in the observer's frame (Rybicki \& Lightman 1979), 
\begin{equation}
\label{dpdw11}
\frac{d P(\nu)}{d \Omega} = \frac{1}{\gamma^3 (1 - \beta \mu)^3}
			    \frac{dP'(\nu ')}{d \Omega '}
                          = \frac{1}{\gamma^3 (1 - \beta \mu)^3}
			    \frac{P'(\nu ')}{4 \pi},
\end{equation}
\begin{equation}
\label{nu12}
\nu = \frac{\nu '}{\gamma (1 - \mu \beta)}. 
\end{equation}
Then the observed flux density at frequency $\nu$ is (Huang et al. 2000a, b),
\begin{equation}
\label{snue27}
S_{\nu} = \frac{1}{A} \left( \frac{dP(\nu)}{d \Omega} \frac{A}{D_{\rm L}^2} \right)
        = \frac{1}{\gamma^3 (1 - \beta \mu)^3} \frac{1}{4 \pi D_{\rm L}^2}
          P'\left(\gamma(1 - \mu \beta) \nu \right),
\end{equation}
where $A$ is the area of our detector and $D_{\rm L}$ is the luminosity 
distance.

\section{Analytic Solutions}

Since the geometry concerned here is still quite simple, we can derive
approximate solutions to the problem analytically. 
In this section, for simplicity, we assume $\Theta = 0$, i.e., the jet
is just moving along the line of sight. Obviously, we then do not need 
to consider the effect of the equal arrival time
surfaces (Waxman 1997b; Sari 1997; Panaitescu \& M\'{e}sz\'{a}ros 1998).
We discuss the two cases of
$v_{\perp} \equiv 0$ and $v_{\perp} \equiv c_{\rm s}$ separately.
Note that we take $\epsilon \equiv 0$ all through the article.

\subsection{$v_{\perp} \equiv 0$ case}

During the ultra-relativistic phase, since $\gamma \gg 1$,
$\beta \approx 1$, and $\gamma m \gg M_{\rm ej}$, equations~(\ref{ndgm2})
--- (\ref{dadt5}) reduce to,
\begin{equation}
\label{diff14}
\frac{d \gamma}{d m} \approx - \frac{\gamma}{2 m}, \;\;
\frac{d m}{d R} \approx {\rm const}, \;\;
\frac{d R}{d t} \approx 2 \gamma^2 c.
\end{equation}
Then it is very easy to get:
\begin{equation}
\label{scale15}
\gamma \propto m^{-1/2}, \;\; m \propto R, \;\; R \propto t^{1/2}.
\end{equation}
Additionally, from the shock conditions (Blandford \& McKee 1976;
Huang et al. 1998a, b), the co-moving thermal energy density is
$e' \propto \gamma^2 \propto t^{-1/2}$, so we have
$B' \propto (e')^{1/2} \propto t^{-1/4}$, and 
$\nu_{\rm c} \propto \gamma \gamma_{\rm c}^2 B' \propto t^{-1}$. 
Since the peak frequency of synchrotron radiation is
$\nu_{\rm m} \propto \gamma \gamma_{\rm e,min}^2 B'$ (Sari et al. 1998),
we have $\nu_{\rm m} \propto t^{-1}$, and the peak flux density is
$S_{\nu,{\rm max}} \propto N_{\rm e} \gamma B' / \gamma^{-2}
\propto N_{\rm e} \gamma^3 B' \propto t^{-1/2}$. The
factor of $\gamma^{-2}$ in $S_{\nu,{\rm max}}$ is due to the
relativistic beaming effect. At last, we can derive the flux density
of the afterglow as follows,
\begin{equation}
\label{snu16}
S_{\nu} \approx \left \{ 
   \begin{array}{lll}
       S_{\nu, {\rm max}} \left( \frac{\nu}{\nu_{\rm m}} \right) ^{-(p-1)/2}
           \propto \nu^{-(p-1)/2} t^{-p/2}, \;\; 
           (\nu_{\rm m} \leq \nu < \nu_{\rm c}, \;\; 
           {\rm slow \;\; cooling} ), \\ \mbox{} \\
       S_{\nu, {\rm max}} \left( \frac{\nu_{\rm c}}{\nu_{\rm m}}\right) 
           ^{-(p-1)/2} \left( \frac{\nu}{\nu_{\rm c}} \right) ^{-p/2}
           \propto \nu^{-p/2} t^{-(p+1)/2}, \;\; 
           (\nu_{\rm m} < \nu_{\rm c} \leq \nu, \;\; 
           {\rm slow \;\; cooling} ), \\ \mbox{} \\
       S_{\nu, {\rm max}} \left( \frac{\nu}{\nu_{\rm c}} \right) ^{-1/2}
           \propto \nu^{-1/2} t^{-1}, \;\; (\nu_{\rm c} \leq \nu < 
           \nu_{\rm m}, \;\; {\rm fast \;\; cooling} ), \\ \mbox{} \\
       S_{\nu, {\rm max}} \left( \frac{\nu_{\rm m}}{\nu_{\rm c}}\right) 
           ^{-1/2} \left( \frac{\nu}{\nu_{\rm m}} \right) ^{-p/2}
           \propto \nu^{-p/2} t^{-(p+1)/2}, \;\; 
           (\nu_{\rm c} < \nu_{\rm m} \leq \nu, \;\; 
           {\rm fast \;\; cooling} ). \\
   \end{array}
   \right.
\end{equation}

In the non-relativistic phase, $\gamma \sim 1$, $\beta \ll 1$,
$d \gamma \approx \beta d \beta$, equations~(\ref{ndgm2}) ---
(\ref{dadt5}) reduce to,
\begin{equation}
\label{diff17}
\frac{d \beta}{d m} \approx - \frac{\beta}{2 m}, \;\;
\frac{d m}{d R} \approx {\rm const}, \;\;
\frac{d R}{d t} \approx \beta c.
\end{equation}
The scaling laws become,
\begin{equation}
\label{scale18}
\beta \propto m^{-1/2}, \;\; m \propto R, \;\; R \propto t^{2/3}.
\end{equation}
According to the shock conditions (Huang et al. 1998a, b), the co-moving
thermal energy density is $e' \propto \gamma-1 \propto t^{-2/3}$.
Then we have $B' \propto (e')^{1/2} \propto t^{-1/3}$,
$\nu_{\rm c} \propto \gamma \gamma_{\rm c}^2 B' \propto t^{-1}$, 
$\nu_{\rm m} \propto \gamma \gamma_{\rm e,min}^2 B' \propto t^{-5/3}$,
$S_{\nu,{\rm max}} \propto \gamma N_{\rm e} B' \propto t^{1/3}$, and
finally the flux density is,
\begin{equation}
\label{snu19}
S_{\nu} \approx \left \{ 
   \begin{array}{lll}
       S_{\nu, {\rm max}} \left( \frac{\nu}{\nu_{\rm m}} \right) ^{-(p-1)/2}
           \propto \nu^{-(p-1)/2} t^{-(5p-7)/6}, \;\; 
           (\nu_{\rm m} \leq \nu < \nu_{\rm c}, \;\; 
           {\rm slow \;\; cooling} ), \\ \mbox{} \\
       S_{\nu, {\rm max}} \left( \frac{\nu_{\rm c}}{\nu_{\rm m}}\right) 
           ^{-(p-1)/2} \left( \frac{\nu}{\nu_{\rm c}} \right) ^{-p/2}
           \propto \nu^{-p/2} t^{-(5p-4)/6}, \;\; 
           (\nu_{\rm m} < \nu_{\rm c} \leq \nu, \;\;
           {\rm slow \;\; cooling} ), \\ \mbox{} \\
       S_{\nu, {\rm max}} \left( \frac{\nu}{\nu_{\rm c}} \right) ^{-1/2}
           \propto \nu^{-1/2} t^{-1/6}, \;\; (\nu_{\rm c} \leq \nu < 
           \nu_{\rm m}, \;\; {\rm fast \;\; cooling} ), \\ \mbox{} \\
       S_{\nu, {\rm max}} \left( \frac{\nu_{\rm m}}{\nu_{\rm c}}\right) 
           ^{-1/2} \left( \frac{\nu}{\nu_{\rm m}} \right) ^{-p/2}
           \propto \nu^{-p/2} t^{-(5p-4)/6}, \;\; 
           (\nu_{\rm c} < \nu_{\rm m} \leq \nu, \;\; 
           {\rm fast \;\; cooling} ). \\
   \end{array}
   \right.
\end{equation}

Let us discuss the decay of optical afterglows (e.g., R band, at the 
frequency $\nu_{\rm R}$) in more detail. In the ultra-relativistic phase, 
$\nu_{\rm m} \ll \nu_{\rm R} \ll \nu_{\rm c}$ is typically satisfied 
initially, then we have $S_{\rm R} \propto \nu^{-(p-1)/2} t^{-p/2}$ at 
early stages. Since $\nu_{\rm m}$ and $\nu_{\rm c}$ both scale as $t^{-1}$, 
we will eventually have $\nu_{\rm m} < \nu_{\rm c} < \nu_{\rm R}$, then
the afterglow at later stages is 
just $S_{\rm R} \propto \nu^{-p/2} t^{-(p+1)/2}$, i.e., the light 
curve steepens by $t^{1/2}$ due to the cooling effect of electrons. 
After entering the non-relativistic phase, 
since $\nu_{\rm m} \propto t^{-5/3}$ and $\nu_{\rm c} \propto t^{-1}$, 
the relation of $\nu_{\rm m} < \nu_{\rm c} < \nu_{\rm R}$ will continue 
to be satisfied. The afterglow then decays 
as $S_{\rm R} \propto \nu^{-p/2} t^{-(5p-4)/6}$, which means the light 
curve usually becomes flatter again, by $t^{(7-2p)/6}$

\subsection{$v_{\perp} \equiv c_{\rm s}$ case}

During the ultra-relativistic phase, $\gamma \gg 1$, $\beta \approx 1$,
$\gamma m \gg M_{\rm ej}$, $c_{\rm s} \approx c/\sqrt{3}$,
equations~(\ref{ndgm2}) --- (\ref{dadt5}) reduce to,
\begin{equation}
\label{diff20}
\frac{d \gamma}{d m} \approx - \frac{\gamma}{2 m}, \;\;
\frac{d m}{d R} \approx \pi a^2 n m_{\rm p}, \;\;
\frac{d a}{d t} \approx \frac{2 \gamma c}{\sqrt{3}}, \;\;
\frac{d R}{d t} \approx 2 \gamma^2 c.
\end{equation}
The corresponding solution is, 
\begin{equation}
\label{scale21}
\gamma \propto t^{-1/2}, \;\; a \propto t^{1/2}, \;\;
      m \propto t.
\end{equation}
Similarly we have $e' \propto \gamma^2 \propto t^{-1}$, 
$B' \propto (e')^{1/2} \propto t^{-1/2}$,
$\nu_{\rm c} \propto \gamma \gamma_{\rm c}^2 B' \propto t^0$,
$\nu_{\rm m} \propto \gamma \gamma_{\rm e,min}^2 B'
\propto t^{-2}$, and
$S_{\nu,{\rm max}} \propto N_{\rm e} \gamma B' / \gamma^{-2}
\propto t^{-1}$.
The flux density of the afterglow is then given by,
\begin{equation}
\label{snu22}
S_{\nu} \approx \left \{ 
   \begin{array}{lll}
       S_{\nu, {\rm max}} \left( \frac{\nu}{\nu_{\rm m}} \right) ^{-(p-1)/2}
           \propto \nu^{-(p-1)/2} t^{-p}, \;\; 
           (\nu_{\rm m} \leq \nu < \nu_{\rm c}, \;\;
           {\rm slow \;\; cooling} ), \\ \mbox{} \\
       S_{\nu, {\rm max}} \left( \frac{\nu_{\rm c}}{\nu_{\rm m}}\right) 
           ^{-(p-1)/2} \left( \frac{\nu}{\nu_{\rm c}} \right) ^{-p/2}
           \propto \nu^{-p/2} t^{-p}, \;\; 
           (\nu_{\rm m} < \nu_{\rm c} \leq \nu, \;\; 
           {\rm slow \;\; cooling} ), \\ \mbox{} \\
       S_{\nu, {\rm max}} \left( \frac{\nu}{\nu_{\rm c}} \right) ^{-1/2}
           \propto \nu^{-1/2} t^{-1}, \;\; (\nu_{\rm c} \leq \nu < 
           \nu_{\rm m}, \;\; {\rm fast \;\; cooling} ) \\ \mbox{} \\
       S_{\nu, {\rm max}} \left( \frac{\nu_{\rm m}}{\nu_{\rm c}}\right) 
           ^{-1/2} \left( \frac{\nu}{\nu_{\rm m}} \right) ^{-p/2}
           \propto \nu^{-p/2} t^{-p}, \;\; 
           (\nu_{\rm c} < \nu_{\rm m} \leq \nu, \;\; 
           {\rm fast \;\; cooling} ). \\
   \end{array}
   \right.
\end{equation}
Note that in most cases, the flux density decays as $t^{-p}$. 

In the non-relativistic phase, $\gamma \sim 1$, $\beta \ll 1$,
$c_{\rm s} \approx \sqrt{5} \beta c/3$, equations~(\ref{ndgm2}) ---
(\ref{dadt5}) reduce to,
\begin{equation}
\label{diff23}
\frac{d \beta}{d m} \approx - \frac{\beta}{2 m}, \;\;
\frac{d m}{d R} \approx \pi a^2 n m_{\rm p}, \;\;
\frac{d a}{d t} \approx \frac{\sqrt{5} \beta c}{3}, \;\;
\frac{d R}{d t} \approx \beta c.
\end{equation}
The solution is also easy to get as,
\begin{equation}
\label{scale24}
\beta \propto m^{-1/2}, \;\; m \propto R^3, \;\; R \propto a \propto t^{2/5}.
\end{equation}
Then we have: $e' \propto \gamma-1 \propto t^{-6/5}$, 
$B' \propto (e')^{1/2} \propto t^{-3/5}$,
$\nu_{\rm c} \propto \gamma \gamma_{\rm c}^2 B' \propto t^{-1/5}$, 
$\nu_{\rm m} \propto \gamma \gamma_{\rm e,min}^2 B' \propto
\beta^4 B' \propto t^{-3}$,
$S_{\nu,{\rm max}} \propto \gamma N_{\rm e} B' \propto t^{3/5}$.
So the flux density is,
\begin{equation}
\label{snu25}
S_{\nu} \approx \left \{ 
   \begin{array}{lll}
       S_{\nu, {\rm max}} \left( \frac{\nu}{\nu_{\rm m}} \right) ^{-(p-1)/2}
           \propto \nu^{-(p-1)/2} t^{-(15p-21)/10}, \;\; 
           (\nu_{\rm m} \leq \nu < \nu_{\rm c}, \;\;
           {\rm slow \;\; cooling} ), \\ \mbox{} \\
       S_{\nu, {\rm max}} \left( \frac{\nu_{\rm c}}{\nu_{\rm m}}\right) 
           ^{-(p-1)/2} \left( \frac{\nu}{\nu_{\rm c}} \right) ^{-p/2}
           \propto \nu^{-p/2} t^{-(15p-20)/10}, \;\; 
           (\nu_{\rm m} < \nu_{\rm c} \leq \nu, \;\; 
           {\rm slow \;\; cooling} ), \\ \mbox{} \\
       S_{\nu, {\rm max}} \left( \frac{\nu}{\nu_{\rm c}} \right) ^{-1/2}
           \propto \nu^{-1/2} t^{1/2}, \;\; (\nu_{\rm c} \leq \nu < 
           \nu_{\rm m}, \;\; {\rm fast \;\; cooling} ), \\ \mbox{} \\
       S_{\nu, {\rm max}} \left( \frac{\nu_{\rm m}}{\nu_{\rm c}}\right) 
           ^{-1/2} \left( \frac{\nu}{\nu_{\rm m}} \right) ^{-p/2}
           \propto \nu^{-p/2} t^{-(15p-20)/10}, \;\; 
           (\nu_{\rm c} < \nu_{\rm m} \leq \nu, \;\; 
           {\rm fast \;\; cooling} ). \\ 
   \end{array}
   \right.
\end{equation}

Again let us have a look at the R band optical light curve. In the 
ultra-relativistic phase, typically we 
have $\nu_{\rm m} \ll \nu_{\rm R} \ll \nu_{\rm c}$, so 
the afterglow decays as $S_{\rm R} \propto \nu^{-(p-1)/2} t^{-p}$. 
After entering the non-relativistic phase, we will
have $\nu_{\rm m} \leq \nu_{\rm R} < \nu_{\rm c}$ at first, which means 
$S_{\rm R} \propto \nu^{-(p-1)/2} t^{-(15p-21)/10}$. We see that 
the light curve becomes flatter by $t^{(21-5p)/10}$. Since $\nu_{\rm c}$
now decreases as $\nu_{\rm c} \propto t^{-1/5}$, at sufficiently  late 
stages, we will eventually have $\nu_{\rm m} < \nu_{\rm c} < \nu_{\rm R}$,
which leads to $S_{\rm R} \propto \nu^{-p/2} t^{-(15p-20)/10}$, i.e.,
the light curve is still flatter than that in the relativistic phase 
(by $t^{(20-5p)/10}$). In short, after entering the non-relativistic phase,
the light curve is expected to become flatter obviously, which is quite 
different from the behaviour of a laterally expanding conical jet (Huang 
et al. 2000a, b, c, d). 

\section{Numerical Results}

Now we solve equations~(\ref{ndgm2}) --- (\ref{dadt5}) numerically and
compare the optical (R band) afterglow light curves with those of conical
jets. First we describe our initial values and parameters briefly. The
ISM density ($n$), the original ejecta mass ($M_{\rm ej}$) and the initial
value of $\gamma$ are fixedly set as $n = 1$ cm$^{-3}$, $M_{\rm ej} =
5 \times 10^{-10} M_{\odot}$, and $\gamma_0 = 300$, respectively. We
further assume $a_0 = 1.2 \times 10^{14}$ cm, then the initial values of
$m$ and $R$ are given by: $m_0 = M_{\rm ej}/300$,
$R_0 = m_0/(\pi a_0^2 n m_{\rm p}) \approx 1.2 \times 10^{17}$ cm.
These values indicate that the angular radius of our GRB ejecta
from the central engine is only $ \sim 10^{-3}$ at the beginning, 
and in the
ultra-relativistic phase considered here it is just as powerful as an
isotropic fireball with kinetic energy $E_{\rm iso,0} \sim 10^{54}$ ergs.
In calculating the R band flux densities,
we take $D_{\rm L} = 1$ Gpc, and will assume $p=2.5$, $\xi_{\rm e} = 0.1$,
$\xi_{\rm B}^2 = 10^{-6}$, $\Theta = 0$ unless declared explicitly.

Here, we would like to point out that many of our initial values and 
parameters (such as $n$, $M_{\rm ej}$, $\gamma_0$, $a_0$ and 
$D_{\rm L}$) have been arbitrarily chosen. However, variation 
of them within reasonable ranges will not alter the major conclusions 
we draw below in this section. 

\subsection{$v_{\perp} \equiv 0$ case}

Assuming no lateral expansion, we have evaluated equations~(\ref{ndgm2})
--- (\ref{dadt5}) numerically. Fig.~1 shows the evolution of the Lorentz
factor, $\gamma$. The jet becomes non-relativistic after $\sim 10^{11}$ s.
In the relativistic phase the slope is approximately $-0.26$ and in the
non-relativistic phase it is about $-0.68$, both consistent with the
theoretical value (i.e., $-1/4$ and $-2/3$ respectively, see Section 3.1).
For comparison, the dashed line shows the evolution of $\gamma$ for a
conical jet (parameters: $E_{\rm iso,0} = 10^{54}$ ergs, $\gamma_0 = 300$,
with an initial half opening angle $\theta_0 = 0.1$). We see that the
cylindrical jet decelerates much slower, which is easy to understand.

Fig.~2 illustrates the Lorentz factor as a function of $R$. The slope
of the solid line is $-1.00$ in the relativistic phase and $-0.50$ in the
non-relativistic phase, also consistent with the theoretical results.
This further confirms the correctness of our programs.

The R band afterglow light curves are shown in Fig.~3. The solid line
represents a cylindrical jet with $p = 2.5$. 
Its slope is approximately $-1.26$
before $10^{9}$ s, consistent with the theoretical value of $-1.25$
(for $\nu_{\rm m} < \nu_{\rm R} < \nu_{\rm c}$ case, see Section 3.1). 
The slope is $-1.89$ between $10^{10}$ s and $10^{11}$ s, also
in good agreement with our analytical results ($-1.75$ for 
$\nu_{\rm m} < \nu_{\rm c} < \nu_{\rm R}$).
After $\sim 10^{11}$ s, the jet enters the non-relativistic phase and 
the light curve becomes flatter slightly, just as required by the 
theoretical result. The dotted line here
represents afterglows from a conical jet ($p=2.5$). 
We can see that the light curve of a cylindrical jet 
differs markedly from that of a conical one, mainly in
the following two aspects: (i) the relativistic phase lasts for a very 
long period ($> 10^{10}$ s typically), during which only a slight break 
(by $t^{1/2}$, due to cooling of electrons) can be observed; 
(ii) after entering the Newtonian phase, it becomes flatter slightly, not
steeper. Please note that at the end point of each line, the characteristic
Lorentz factor of electrons has already been as small as $\gamma_{\rm e}
\sim 5$.

Fig.~4 illustrates the effect of three parameters, $\xi_{\rm e}$, 
$\xi_{\rm B}^2$ and $\Theta$, 
on the light curve. It is clear that $\xi_{\rm e}$ 
does not change the shape of the light curve markedly. However, 
because $\nu_{\rm c}$ strongly depends on $B'$ ($\gamma_{\rm c} 
\propto B'^{-2}$, $\nu_{\rm c} \propto B'^{-3}$), we see that 
$\xi_{\rm B}^2$  affects the cooling break seriously. 
In our calculations, since we take $\gamma_0 = 300$, we have 
found that for any $\Theta$ smaller than 1/300, the light curve 
will be almost identical to the $\Theta = 0$ case. But in 
the $\Theta = 0.1$ case, the flux density is lowered down 
markedly before $\sim 10^7$ s, which means a cylindrical jet is 
observable only within a narrow solid angle. 

\subsection{$v_{\perp} \equiv c_{\rm s}$ case}

When lateral expansion is included, the results are quite different.
The solid line in Fig.~5 illustrates the evolution of $\gamma$ in
this case. We can see that the jet becomes 
Newtonian at $\sim 10^5$ --- $10^6$ s,
much earlier than the $v_{\perp} \equiv 0$ case. Note that its slope
is approximately $-0.51$ in the ultra-relativistic phase and $-1.20$
in the non-relativistic phase, both consistent with the theoretical
values (i.e., $-1/2$ and $-6/5$ respectively, see Section 3.2). Fig.~5
also shows that when $v_{\perp} \equiv c_{\rm s}$, a cylindrical jet
decelerates faster than a conical one, which is just contrary to the
$v_{\perp} \equiv 0$ case. The $\gamma$ -- $R$ relation is shown in
Fig.~6 by the solid line. In the non-relativistic phase, the slope
is $\sim -3.0$, consistent with the theoretical value. However, in
the ultra-relativistic phase, the $\gamma$ -- $R$ relation is not a
power-law function, but an exponential one.

Fig.~7 illustrates the evolution of the jet profile on the $y$ -- $z$
plane. Generally speaking, the lateral expansion is not significant
during the ultra-relativistic phase. But in the non-relativistic phase,
the lateral radius increases synchronically with the radial scale.

R band afterglows from cylindrical jets expanding laterally are shown in
Fig.~8. The solid line is plot with $p=2.5$. Its slope is $\sim -2.63$
before $10^6$ s, consistent with the theoretical value of $-2.5$
(see Section 3.2). It turns flatter in the non-relativistic phase, also
consistent with our analytical results. The dotted line corresponds
to a laterally expanding conical jet ($p=2.5$), where a break due to the
relativistic-Newtonian transition can be clearly seen (Huang 
et al. 2000a, b, c, d). It is clear that the light curve of a
cylindrical jet differs from that of a conical one mainly in the following
aspects: (i) no obvious breaks could be observed; (ii) in the
relativistic phase, the decay is always as quick as $t^{-p}$, but
for a conical jet the decay is only $t^{-(3p-3)/4}$ at 
early stages (Rhoads 1997, 1999);
(iii) after entering the non-relativistic phase, the light curve becomes
flatter slightly, also quite different from that of a conical jet.
Fig.~9 illustrates the effect of $\xi_{\rm e}$, $\xi_{\rm B}^2$
and $\Theta$ on the light curve. Similar to the $v_{\perp} \equiv 0$
case, the flux density is markedly lowered down by a non-zero 
$\Theta$, but generally speaking, $\xi_{\rm e}$ and $\xi_{\rm B}^2$
do not change the slope of the light curve here.

\section{Discussion and Conclusions}

The importance of collimation in GRBs goes beyond influencing the detailed
shape of the light curve: it reveals something about the central engine,
and affects our estimates of how many GRB progenitors there need to be
(van Paradijs et al. 2000; Wijers et al. 1998; Lee, Brown \& Wijers 2000). 
Most theorists now expect GRBs to be
significantly collimated, because jets seem to be produced by all systems
where matter is undergoing disk accretion onto a compact central object,
including young protostars, ``microquasars'' (i.e., black hole binaries in
the Galaxy), capture of a main sequence star by a moderate massive black 
hole in normal galaxies (Cheng \& Lu 2000),  
radio galaxies, and active galactic nuclei (Lamb 2001).
Observations on these systems have interestingly indicated that an
astrophysical jet usually maintains a constant cross section at large
scales (Perley et al. 1984; Biretta et al. 1999). In fact, it has already
been pointed out by Dar et al. (Shaviv \& Dar 1995; Dar 1998, 1999) that
although a conical beam can solve the ``energy crisis'' of GRBs, it
suffers from some other deficiencies of isotropic fireballs, while a
cylindrical jet does not. However, nearly all previous discussion on
afterglows from collimated GRB remnants have assumed a conical geometry,
which, we believe is not enough. Here we have discussed afterglows from
cylindrical jets in detail. Both analytic and numerical results are
presented.

For a cylindrical jet without lateral expansion, 
in the relativistic phase the flux density typically decays
as $S_{\nu} \propto t^{-p/2}$ at first. Then the light curve steepens 
by $t^{1/2}$ due to cooling of electrons. 
After entering the Newtonian phase, the decline becomes slightly slower,
$S_{\nu} \propto t^{-(5p-4)/6}$, just contrary to the behavior of a conical
jet. Note that the remnant usually decelerates very slowly in this case,
i.e., $\gamma \propto t^{-1/4}$ when $\gamma \gg 1$, so that it keeps to be
highly relativistic as long as $t \sim 10^8$ --- $10^9$ s. When lateral
expansion is included and assuming $v_{\perp} \equiv c_{\rm s}$, the
afterglow light curve is $S_{\nu} \propto t^{-p}$ in the ultra-relativistic
phase, and it flattens to be $S_{\nu} \propto t^{-(15p-21)/10}$ or 
$S_{\nu} \propto t^{-(15p-20)/10}$ in the
the non-relativistic phase, also quite different to a conical jet.

In previous sections, we have assumed that the lateral expansion velocity
of our cylindrical jet can be either $v_{\perp} \equiv 0$ or $v_{\perp}
\equiv c_{\rm s}$ and we have not given a preference for either of these
two cases. Now we discuss this question. At the first glance, the
$v_{\perp} \equiv c_{\rm s}$ seems more realistic. However, observations
on relativistic outflows in radio galaxies reveal that lateral expansion
is usually unobtrusive (Perley et al. 1984; Biretta et al. 1999). For
example, observed profiles are generally quite different from the profile
in our Fig.~7. The high degree of collimation 
and the suppression of lateral
expansion is probably due to magnetic confinement (Dar 1998;
Mirabel \& Rodriguez 1999) and/or the external pressure. We believe that
in many cases, even if the cylindrical jet expands laterally, its
velocity will be $v_{\perp} \ll c_{\rm s}$.

Afterglows from some GRBs, such as GRBs 970508, 971214, 980329 and
980703 have been observed to decay slowly and steadily within a long
period (i.e., up to several months, see van Paradijs 
et al. 2000), with $S_{\nu} \propto t^{-1.1}$
--- $t^{-1.3}$. Although an isotropic fireball is the most natural and
popular model for these GRBs, Huang et al. (1998a; also see Dai et al.
1999) pointed out that there is a problem: usually a typical fireball
with kinetic energy $E_{\rm iso,0} = 10^{52}$ ergs will become
non-relativistic in one or two months, then the theoretical light curve
will become steeper slightly (Wijers et al. 1997; 
Huang et al. 1998b, 1999a, b). So it is difficult to
understand why the light curve is a single straight line that lasts
for 0.5 --- 1 year (e.g., GRB 970508). 
Here we suggest that this kind of GRBs might be due
to cylindrical jets (with $v_{\perp} \ll c_{\rm s}$): a light curve with
slope $-1.1$ --- $-1.3$ can be easily produced by assuming
$2.2 \leq p \leq 2.6$, and the steady decline is explained by
the fact that these cylindrical jets are highly relativistic even when
$t \sim 10^8$ --- $10^9$ s. Note that in this model, $\gamma$-rays
of the initial burst are beamed into a solid angle with the angular
radius $\sim 1/\gamma_0$. For $\gamma_0 \sim 300$, the beaming angle
is $\sim 0.003$; and for $\gamma_0 \sim 100$, it is $\sim 0.01$, which
means the energy involved will be greatly reduced.

GRBs 990123, 990510, 991216 and 000301c are characterized by an obvious
break in the afterglow light curve. A break
steeper than $t^{1/2}$ could not be explained directly
by our simple cylindrical jet model, either $v_{\perp} \equiv 0$ or
$v_{\perp} \equiv c_{\rm s}$. However, there are many other factors that
have not been taken into account in this article, such as 
the inhomogeneity of ISM (especially
$n(R) \propto R^{-2}$, Dai \& Lu 1998; Chevalier \& Li 2000), and the
nature of the lateral expansion. For example, we can imagine that in 
some cases, since the shock is highly radiative at first (which means 
the remnant is relatively cool), the magnetic field is much stronger 
initially, and at the same time the external pressure may be high, 
then the lateral expansion can be successfully suppressed, which will
lead to a flat light curve of $S_{\nu} \propto t^{-1}$ --- $t^{-1.3}$ 
at early stages. At later stages the lateral expansion takes effect 
and the light curve becomes $S_{\nu} \propto t^{-2}$ --- $t^{-3}$. 
So an obvious break appears. In short, we suggest that the cylindrical
jet model should be paid attention to and should 
be investigated further since it
seems to be more realistic than the conical jet model on the observational
basis.

Recently, Wang \& Loeb (2001) discussed the emission from bow shocks of 
beamed GRBs. They found that the emitted flux from this bow shock may 
dominate over the direct emission from the conical jet for lines of sight 
which are outside the angular radius of the jet emission. The event rate 
for these lines of sight is thus greatly increased. For typical GRB 
parameters, they found that the bow shock emission from a jet of half-angle 
$\sim 5^{\rm o}$ is visible out to tens of Mpc in the radio and hundreds of 
Mpc in the X-rays. In most of our calculations where
$\Theta = 0$ is assumed, the bow shock emission generally will not play 
a dominant role. However, we should bear in mind that the bow shock emission
is obviously more important for a cylindrical jet than for a conventional 
conical jet, since observers of a cylindrical jet are more likely to be 
situated outside the initial emission cone. For example the bow shock 
emission may take effect in those $\Theta = 0.1$ cases in Figs. 4 and 9. 
A direct effect of bow shock emission from cylindrical jets is that the 
beaming factor will be considerably reduced, so that it will be much easier 
for us to observe them. We will not go further to calculate the bow shock 
emission here. But in the future, if the 
so called ``burstless afterglows'' are observed, then the question of 
bow shock emission will become very important. 

Finally we would like to say a few words about the origin of the 
cylindrical jet. It should not be born a cylinder. In fact, the outflow 
in a black hole-accreting disk system is likely to be conical 
{\em initially}, with the half opening angle $\theta \ll 1$. Assuming 
$\theta \sim 0.1$, then after propagating to a distance of 
$R \sim 10^{15}$ cm, it will acquire a lateral radius of 
$a \sim 10^{14}$ cm. So in our calculation we have implicitly assumed 
that the jet has become cylindrical before $R \sim 10^{15}$ --- $10^{16}$ 
cm. We will not go to any further details of the mechanism that makes 
a conical jet into a cylindrical one here, they are beyond the 
purpose of this article.  However, if our cylindrical jet is really 
originated from a conical outflow very near to the central black 
hole, then we can draw an interesting conclusion as follows. 
Adopting the parameters used in Section 4, our cylindrical jet has 
a total kinetic energy of $E_{\rm ej} \equiv \gamma_0 M_{\rm ej} c^2 
\approx 3 \times 10^{47}$ ergs. Since this energy comes from a cone 
with $\theta \sim 0.1$ when $R \leq 10^{15}$ cm, we can get the 
corresponding $initial$ isotropic energy easily as 
$E_{\rm iso,ini} \equiv 
4 \pi \times E_{\rm ej} / [2 \pi (1 - \cos \theta)] 
\approx 1.2 \times 10^{50}$ ergs. We have already 
pointed out in Section 4 that our cylindrical jet looks just 
as powerful as an isotropic fireball with total kinetic energy 
$E_{\rm iso,0} \sim 10^{54}$ ergs at the GRB afterglow stage 
(i.e., when $R \geq 10^{17}$ cm). So our final conclusion is:
if GRBs are really due to cylindrical jets, then an apparently 
energetic GRB event (eg., $E_{\rm iso} \sim 10^{54}$ ergs ) 
can be easily produced by a supernova-like event (i.e., 
$E_{\rm iso} \sim 10^{50}$ ergs). We even do not need to 
make the embarrassing assumption that all the kinetic energy 
involved in the supernova-like event is beamed into a narrow 
cone to produce the GRB, which, as noted already, is not 
natural. Of course, in this case, the GRB rate in the 
Universe will be much higher correspondingly.

\section*{Acknowledgments}
We are very grateful to R.A.M.J. Wijers for his valuable comments  
and suggestions that lead to an overall improvement 
of this manuscript. 
We also thank F. Yuan, Z. G. Dai, L. Zhang, J. H. Fan and X. Zhang for
stimulative discussion. KSC is supported by a RGC grant of the Hong Kong 
Government, YFH and TL are supported by the National Natural Science 
Foundation of China and the Special Funds for Major State Basic 
Research Projects.

{}

\clearpage

\begin{figure} \centering 
\epsfig{file=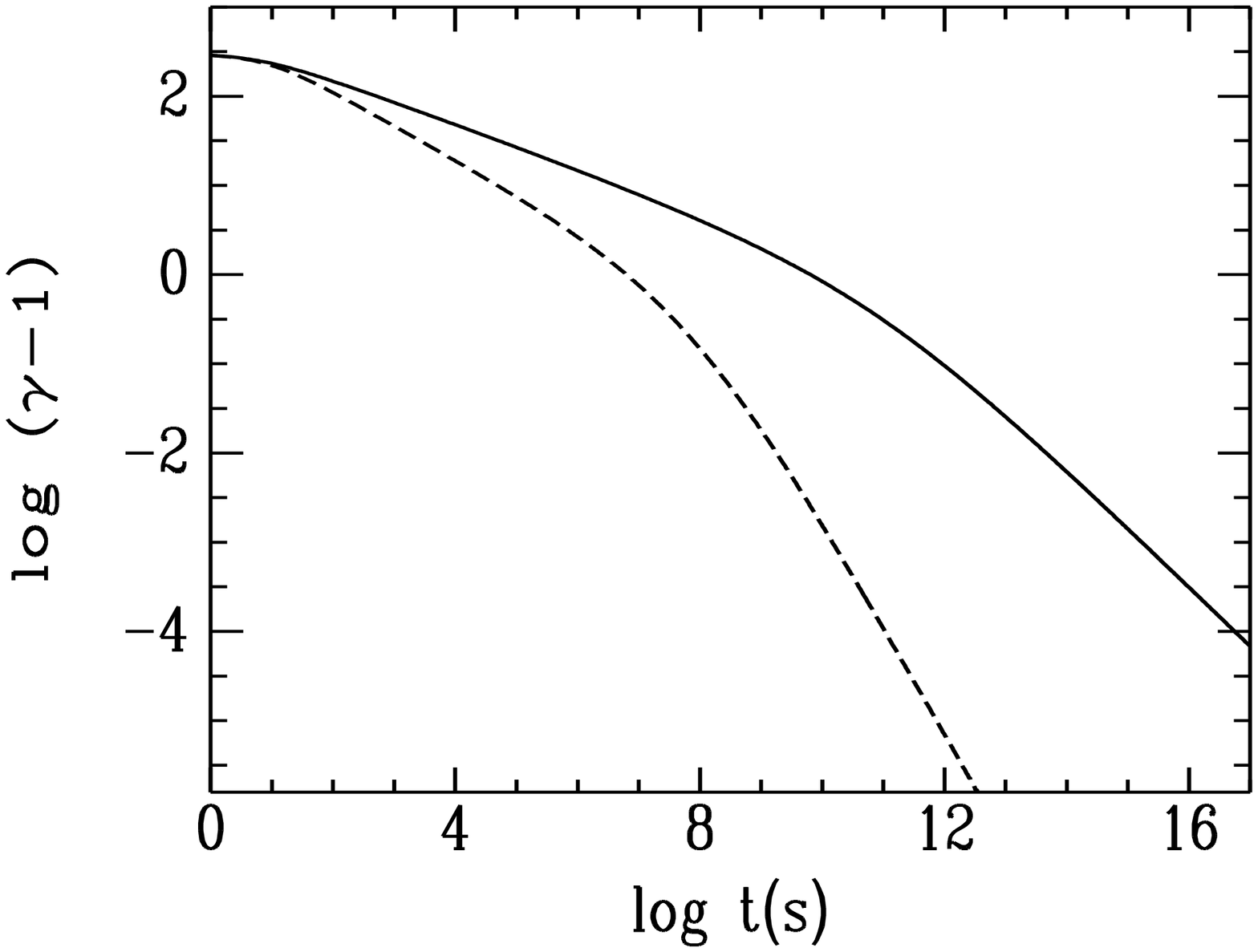, angle=-90, height=60mm, width=6.8cm, 
bbllx=120pt, bblly=125pt, bburx=530pt, bbury=575pt}
\caption{Evolution of the Lorentz factor, $\gamma$, when $v_{\perp}
  \equiv 0$. The solid line corresponds to a cylindrical jet and the
  dashed
  line corresponds to a conical one. Parameters concerned here have been
  described in Section 4 of the main text.}
\label{fig1}
\end{figure}

\begin{figure} \centering
\epsfig{file=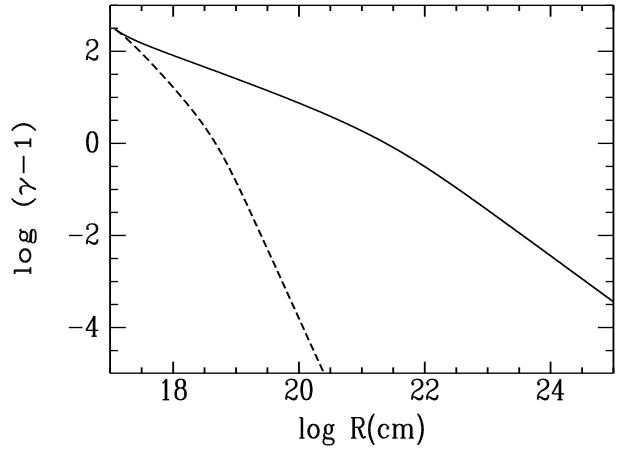, angle=-90, height=60mm, width=6.8cm, 
bbllx=120pt, bblly=125pt, bburx=530pt, bbury=575pt}
\caption{ $\log (\gamma -1)$ vs. $\log R$ for jets without lateral
  expansion ($v_{\perp} \equiv 0$). Line styles and parameters are the
  same as in Fig.~1.}
\label{fig2}
\end{figure}

\begin{figure} \centering 
\epsfig{file=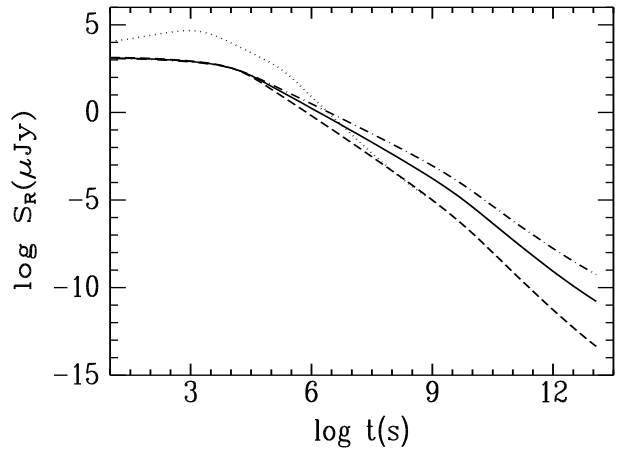, angle=-90, height=60mm, width=6.8cm, 
bbllx=120pt, bblly=125pt, bburx=530pt, bbury=575pt}
\caption{R band afterglows from beamed GRB ejecta without lateral expansion
  ($v_{\perp} \equiv 0$). The dotted line corresponds to a conical jet
  with $p=2.5$. Other lines are for cylindrical jets which differ only in
  the parameter $p$: the dashed, solid and dash-dotted line corresponds
  to $p = 3$, 2.5 and 2.2 respectively. Other parameters have been given
  in Section 4 of the main text. The breaks at $t \sim 10^9$ s in the 
  light curves for cylindrical jets are due to cooling of electrons.}
\label{fig3}
\end{figure}

\begin{figure} \centering
\epsfig{file=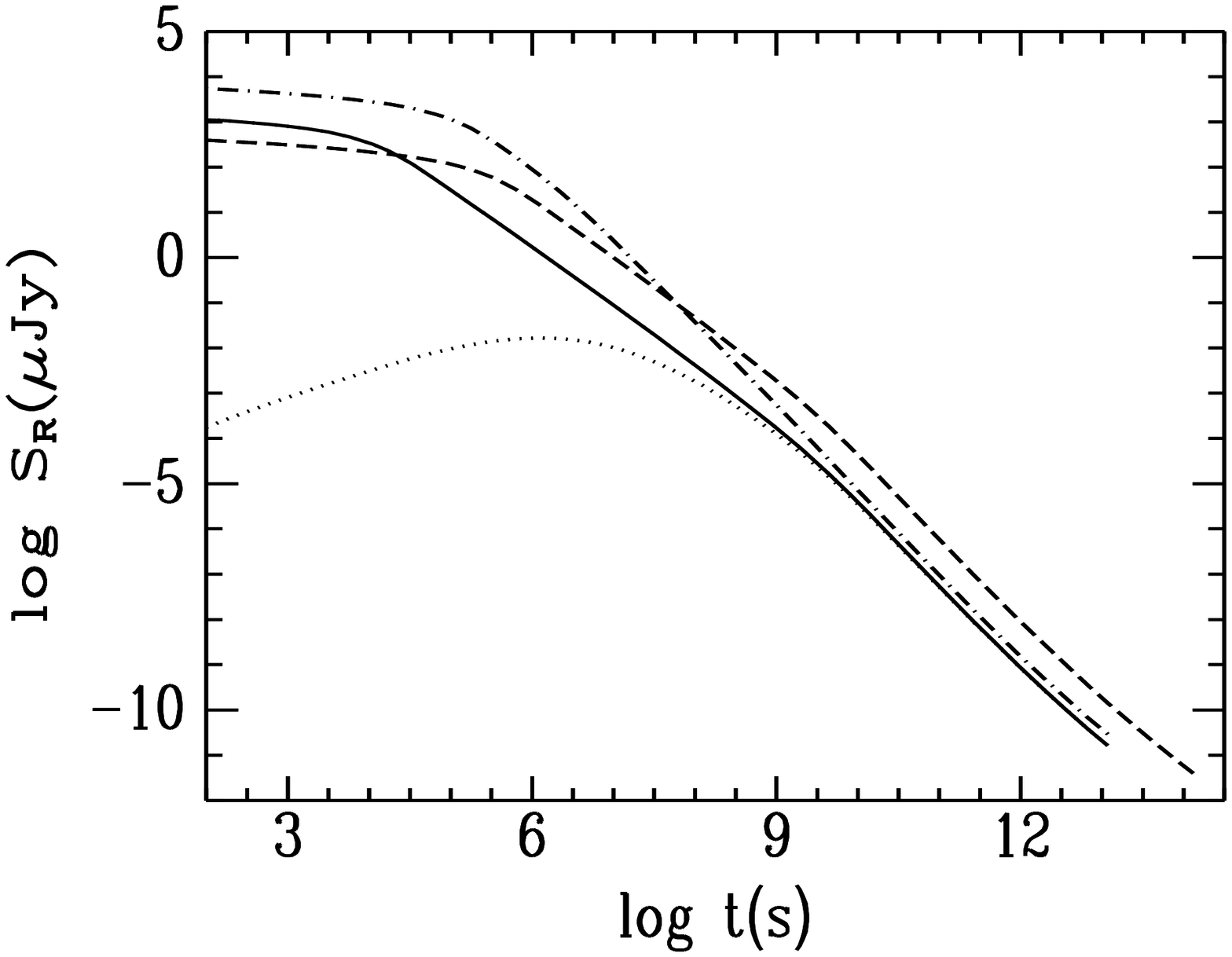, angle=-90, height=60mm, width=6.8cm, 
bbllx=120pt, bblly=125pt, bburx=530pt, bbury=575pt}
\caption{The effect of $\xi_{\rm e}$, $\xi_{\rm B}^2$ 
   and $\Theta$ on the R band
   light curve ($v_{\perp} \equiv 0$ case). The solid line is plot with
   $\xi_{\rm e} = 0.1$, $\xi_{\rm B}^2 = 10^{-6}$, and $\Theta = 0$. 
   Each of the other lines is plot with only one parameter altered 
   with respect to the solid line. The dashed, dash-dotted and dotted lines
   correspond to $\xi_{\rm e} = 0.5$, $\xi_{\rm B}^2 =10^{-4}$ and 
   $\Theta = 0.1$ respectively. Other parameters not mentioned here
   have been described in Section 4 of the main text. }
\label{fig4}
\end{figure}

\begin{figure} \centering 
\epsfig{file=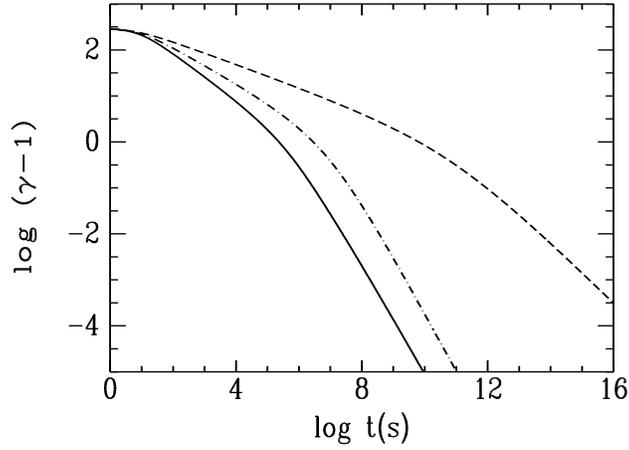, angle=-90, height=60mm, width=6.8cm, 
bbllx=120pt, bblly=125pt, bburx=530pt, bbury=575pt}
\caption{Evolution of the Lorentz factor, $\gamma$. The solid line
  corresponds to a cylindrical jet with $v_{\perp} \equiv c_{\rm s}$.
  For comparison, the dash-dotted line corresponds to a conical jet
  with $v_{\perp} \equiv c_{\rm s}$, and the dashed line is for a
  cylindrical jet without lateral expansion. Parameters are the same
  as in Fig. 1.}
\label{fig5}
\end{figure}

\begin{figure} \centering
\epsfig{file=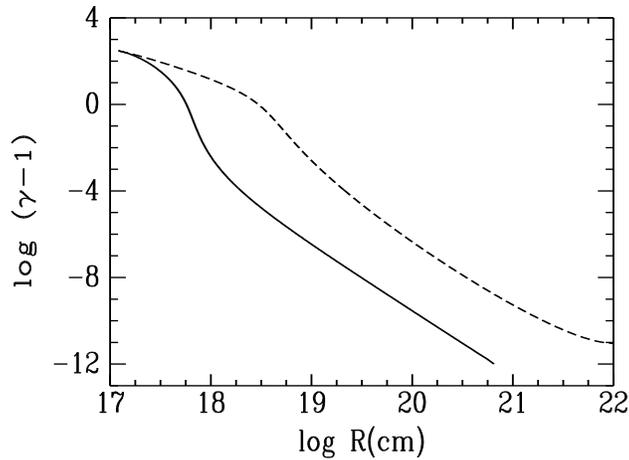, angle=-90, height=60mm, width=6.8cm, 
bbllx=120pt, bblly=125pt, bburx=530pt, bbury=575pt}
\caption{$\log (\gamma -1)$ vs. $\log R$ for jets with lateral expansion
 $v_{\perp} \equiv c_{\rm s}$. The solid line corresponds to a cylindrical
 jet and the dashed line corresponds to a conical one. Parameters
 are the same as in Fig. 1.}
\label{fig6}
\end{figure}

\begin{figure} \centering 
\epsfig{file=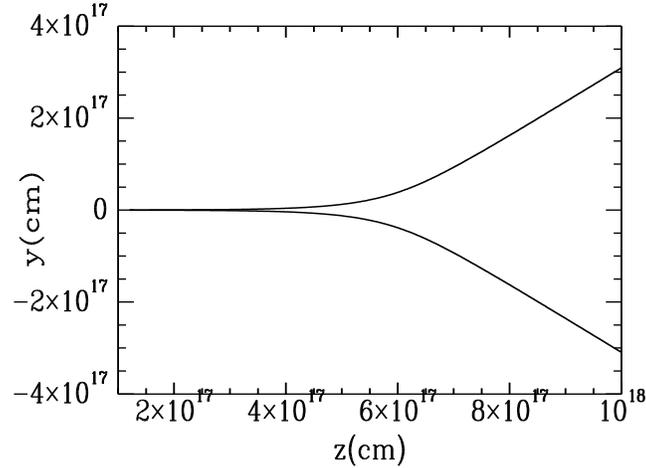, angle=-90, height=60mm, width=6.8cm, 
bbllx=120pt, bblly=125pt, bburx=530pt, bbury=575pt}
\caption{Schematic illustration of the lateral expansion of a
   cylindrical
   jet on the $y$ -- $z$ plane. $z$-axis is the symmetry axis of the jet
   and the lateral expansion is along $y$-axis. Parameters are the same
   as those associated with the solid line in Fig. 6.}
\label{fig7}
\end{figure}

\begin{figure} \centering 
\epsfig{file=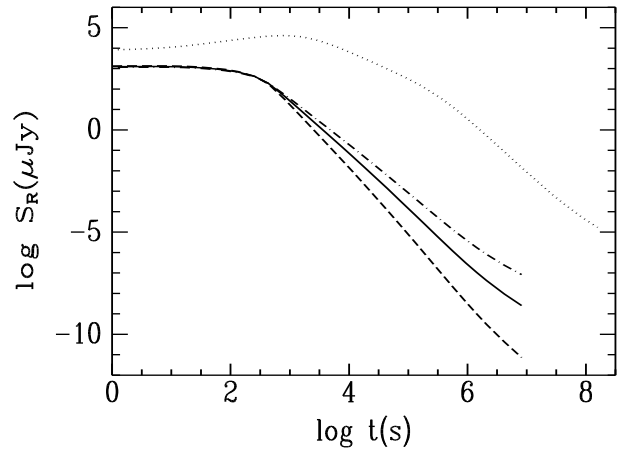, angle=-90, height=60mm, width=6.8cm, 
bbllx=120pt, bblly=125pt, bburx=530pt, bbury=575pt}
\caption{R band afterglows from beamed GRB ejecta with lateral expansion
  ($v_{\perp} \equiv c_{\rm s}$). The dotted line corresponds to a
  conical jet with $p = 2.5$, other lines are for cylindrical jets which
  differ only in the parameter $p$. The dash-dotted, solid and the dashed 
  line corresponds to $p = 2.2$, 2.5 and 3.0 respectively. Other
  parameters not mentioned here have been described in Section 4 of the
  main text. Note that the cylindrical jets here are already 
  non-relativistic when $t > 10^6$ s.}
\label{fig8}
\end{figure}

\begin{figure} \centering 
\epsfig{file=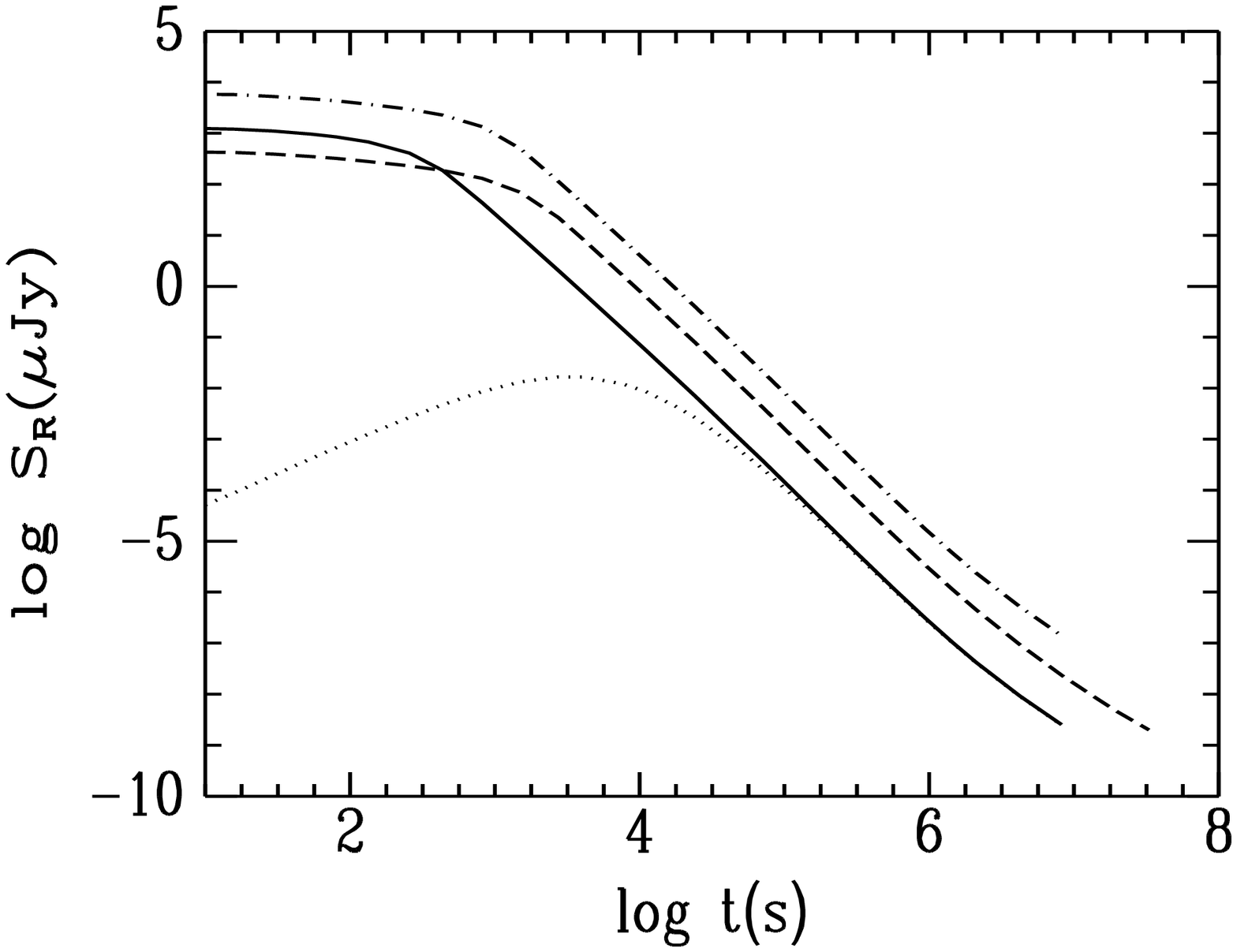, angle=-90, height=60mm, width=6.8cm, 
bbllx=120pt, bblly=125pt, bburx=530pt, bbury=575pt}
\caption[]{The effect of $\xi_{\rm e}$, $\xi_{\rm B}^2$ and 
   $\Theta$ on the R band
   light curve ($v_{\perp} \equiv c_{\rm s}$ case).
   The solid line is plot with
   $\xi_{\rm e} = 0.1$, $\xi_{\rm B}^2 = 10^{-6}$ and $\Theta = 0$. 
   Each of the other lines is plot with only one parameter altered 
   with respect to the solid line.  
   The dashed, dash-dotted and dotted lines
   correspond to $\xi_{\rm e} = 0.5$, $\xi_{\rm B}^2 =10^{-4}$ and 
   $\Theta = 0.1$ respectively. Other parameters not mentioned here
   have been described in Section 4 of the main text.}
\label{fig9}
\end{figure}

\end{document}